\documentclass[12pt,preprint]{aastex}
\usepackage{natbib}
\begin{document}

\title{\sc Discovery of an Unusually Red L-type Brown Dwarf }

\author{John E.\ Gizis,\altaffilmark{1}
Jacqueline K.\ Faherty,\altaffilmark{2}
Michael C. Liu,\altaffilmark{3,4}
Philip J. Castro,\altaffilmark{1}
John D. Shaw,\altaffilmark{5}
Frederick J. Vrba,\altaffilmark{6}
Hugh C. Harris,\altaffilmark{6}
Kimberly M. Aller,\altaffilmark{3,4}
Niall R. Deacon\altaffilmark{4,7}
}

\altaffiltext{1}{Department of Physics and Astronomy, University of Delaware, 
Newark, DE 19716, USA}
\altaffiltext{2}{Department of Astronomy,Universidad de Chile, Cerro Calan, Camino El Observatorio \# 1515, Las Condes Chile}
\altaffiltext{3}{Institute for Astronomy, University of Hawaii, 2680 Woodlawn Drive, Honolulu HI 96822}
\altaffiltext{4}{Visiting Astronomer at the Infrared Telescope Facility, which is operated by the University of Hawaii under Cooperative Agreement no. NNX-08AE38A with the National Aeronautics and Space Administration, Science Mission Directorate, Planetary Astronomy Program.}
\altaffiltext{5}{Department of Physics, West Chester University, West Chester, PA 19383}
\altaffiltext{6}{US Naval Observatory, Flagstaff Station, 10391 West Naval Observatory Road, Flagstaff, AZ 86001}
\altaffiltext{7}{Max-Planck-Institut fur Astronomie, Konigstuhl 17 D-69117, Heidelberg, Germany}

\begin{abstract}
We report the discovery of an unusually red brown dwarf found in a search for high proper motion objects using WISE and 2MASS data.  WISEP J004701.06+680352.1 is moving at $0\farcs44$ yr$^{-1}$ and lies relatively close to the Galactic Plane ($b=5.2^\circ$). Near-infrared photometry and spectroscopy reveals that this is one of the reddest (2MASS J-K$_s = 2.55 \pm 0.08$ mag) field L dwarfs yet detected, making this object an important member of the class of unusually red L dwarfs. We discuss evidence for thick condensate clouds and speculate on the age of the object. Although models by different research groups agree that thick clouds can explain the red spectrum, they predict dramatically different effective temperatures, ranging from 1100K to 1600K. This brown dwarf is well suited for additional studies of extremely dusty substellar atmospheres because it is relatively bright ($K_s = 13.05 \pm 0.03$ mag), which should also contribute to an improved understanding of young gas-giant planets and the transition between L and T brown dwarfs. 
\end{abstract}


\section{Introduction\label{intro}}

Discoveries made with infrared sky surveys have lead to the development of the L \citep{1999AJ....118.2466M,1999ApJ...519..802K}, T \citep{2002ApJ...564..421B,2002ApJ...564..466G} and Y dwarf \citep{Cushing:2011fk} classification systems.  Hundreds of field brown dwarfs have been classified on these systems using optical or near-infrared spectra.  

One of the great benefits of these classification systems is that they allow peculiar objects to be identified. It is possible to identify metal-poor L subdwarfs \citep{2003ApJ...592.1186B,2008ApJ...672.1159B,2009ApJ...694L.140S,2009ApJ...696..986C} as well as very young, low-surface-gravity field brown dwarfs \citep{McGovern:2004dq,2008ApJ...689.1295K,2009AJ....137.3345C} using their spectral features.  Other photometric and spectroscopic peculiarities appear to be associated with the properties of clouds.  Unusually blue L (UBL) dwarfs \citep{2003AJ....126.2421C,2004AJ....127.3553K,2006AJ....131.2722C,Bowler:2010lr} are a few tenths of magnitude bluer in J-K$_s$ color than normal L dwarfs. \citet{2008ApJ...674..451B} argue that they are associated with ``thinner and/or large-grained condensate clouds."  Studies of a binary system consisting of an UBL dwarf
\object{SDSS J141624.08+134826.7} and an unusually blue T dwarf
\object{ULAS J141623.94+134836.3} suggest that system parameters such as old-age/high-gravity and/or low metallicity are responsible for the thin clouds \citep{Bowler:2010lr,2010MNRAS.404.1952B,2010AJ....139.1045S,2010A&A...510L...8S,2010AJ....140.1428C,2010AJ....139.2448B}. \citet{Dupuy:2012lr} argue that their precise parallax of this system favors low metallicity.   

In turn, unusually red L (URL) dwarfs may be associated with thicker clouds \citep{Cushing:2008kx}, caused in turn by low surface gravity and/or high metallicity \citep{McLean:2003lr,2007ApJ...655.1079L,Looper:2008lr,Stephens:2009qy}. \citet{Allers:2010lr} have resolved a binary system which consists of two young URL dwarfs. Kinematics of the two populations support an older age for the UBL dwarfs and a younger age for the URL dwarfs \citep{2009AJ....137....1F,2010AJ....139.1808S}, but there is evidence that the URL population is a mix of ages and includes some relatively old objects  \citep{Looper:2008lr,2010ApJS..190..100K}. Some hot exoplanets and planetary-mass brown dwarf companions, such as \object{HR 8799b} and \object{HR 8799d} \citep{Marois:2008ul,Bowler:2010lp,Barman:2011fk} and \object{2M1207b} \citep{2004A&A...425L..29C,edgeon1207,2010A&A...517A..76P} may be similar to URL dwarfs.

The Wide-field Infrared Survey Explorer (WISE, \citealt{2010AJ....140.1868W}) has now surveyed the entire sky in four mid-infrared filters, with $\sim 58\%$ of the sky included in the April 2011 Preliminary Data Release. The WISE science team \citep{Kirkpatrick:2011lr} and independent groups \citep{Gizis:2011lr,Scholz:2011kx,Loutrel:2011fk,Liu:2011lr} have identified previously overlooked, nearby ultracool dwarfs.  In this paper, we present a new brown dwarf discovered using WISE and the Two Micron All-Sky Survey (2MASS, \citealt{2mass}).  We present the observational data that shows WISEP J004701.06+680352.1 is an unusually red L dwarf (Section~\ref{sec-data}) and compare it to other URL dwarfs and theoretical synthetic spectra (Section~\ref{sec-discussion}).  

\section{Data and Observations\label{sec-data}}

\subsection{WISE/2MASS Selection and Photometry}

\begin{figure}
\plotone{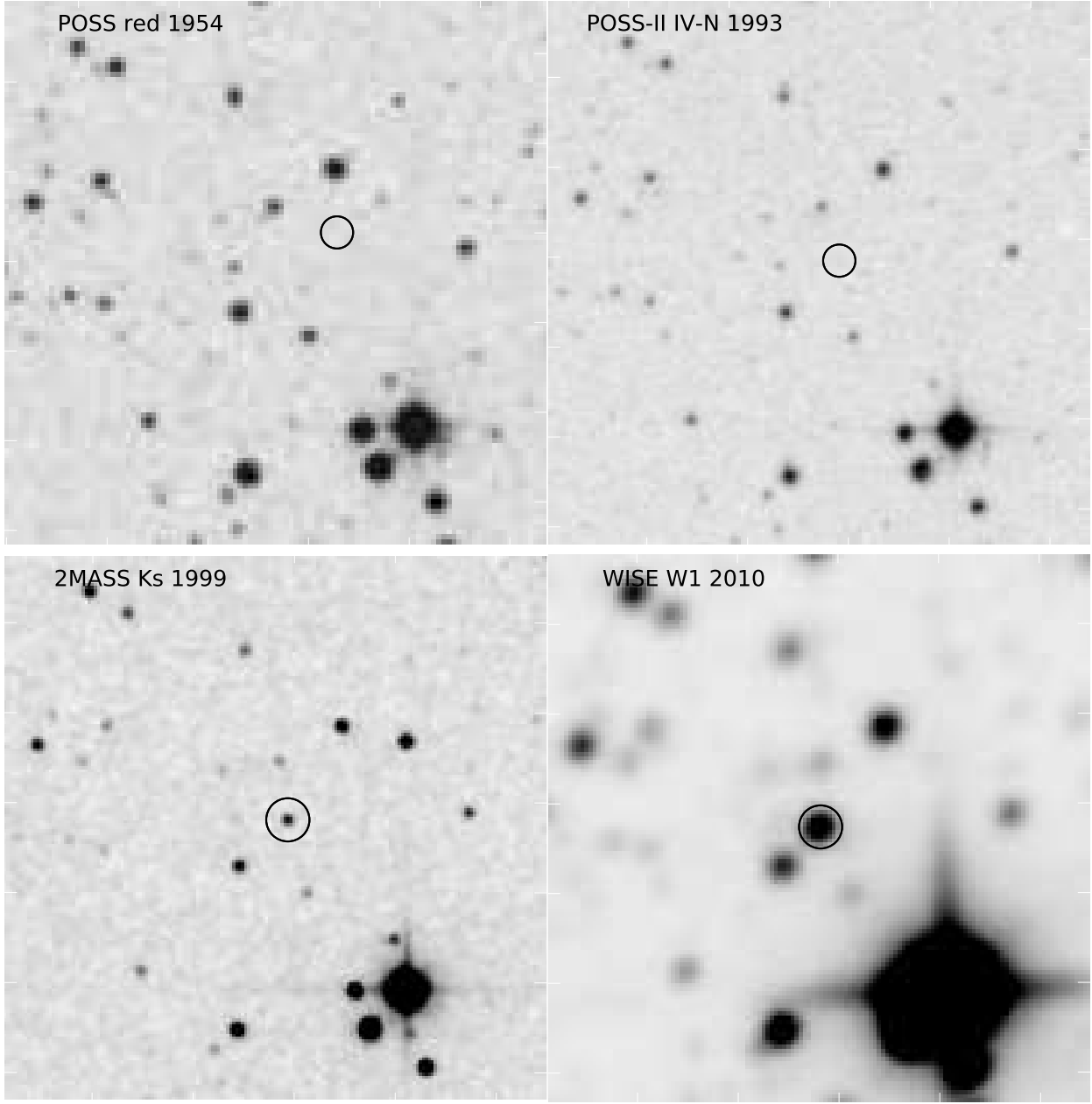}
\caption{Finder charts for W0047+6803. The 2010 WISE W1 \citep{2010AJ....140.1868W}, 1999 2MASS K$_s$ \citep{2mass}, 1993 POSS-II \citep{Reid:1991rt}, and 1954 POSS \citep{Abell:1959fk} images are shown. The circles indicate the position of W0047+6803 at each epoch; it is undetected on the photographic plates.  
\label{fig-finder}}
\end{figure}

WISEP J004701.06+680352.1, hereafter W0047+6803, was identified as a candidate bright (W1$=11.90 \pm 0.02$) L dwarf as part of our search for high proper motion objects (see \citealt{Gizis:2011lr}.) The WISE processing provides an automated pairing with 2MASS sources within 3 arcseconds.  Since the time elapsed between surveys is a decade, objects with motions $\gtrsim 0.3$ arcsec yr$^{-1}$ appear as WISE sources without 2MASS counterparts.  W0047+6803 was selected in a search of such sources with W1$<12$, W1-W2$>0.3$ and low galactic latitude $|b|<10$. Each such source was examined in WISE, 2MASS, and Digitized Sky Survey (DSS) to identify proper motion counterparts. W0047+6803 matches the red 2MASS source 2MASS J00470038+6803543 (K$_s = 13.05 \pm 0.03$) but does not appear on the DSS scans of Palomar photographic plates (Figure~\ref{fig-finder}), suggesting $R_C>20.8$ and $I_C>19.5$ \citep{Reid:1991rt}.  The WISE and 2MASS photometry are listed in Table~\ref{tab1}.   

\subsection{Near-Infrared Spectroscopy}

We obtained low-resolution ($R\approx$150) spectra of W0047+6803 on
21~July~2011~UT from NASA's Infrared Telescope Facility (IRTF) located
on Mauna Kea, Hawaii. Conditions were photometric with typical seeing
conditions ($\approx$0.8\arcsec\ FWHM). We used the near-IR spectrograph
SpeX \citep{2003PASP..115..362R} in prism mode, obtaining
0.8--2.5~\micron\ spectra in a single order. We used the 0.5\arcsec\
wide slit oriented at the parallactic angle to minimize the effect of
atmospheric dispersion. W0047+6803 was nodded along the slit in an ABBA
pattern for a total on-source integration of 8~min, with
individual exposure times of 60~sec. We observed the A0V star \object{HD 12365}
contemporaneously with W0047+6803 for telluric calibration. All spectra
were reduced using version~3.4 of the SpeXtool software package
\citep{2003PASP..115..389V,2004PASP..116..362C}. The S/N per pixel in
the final reduced spectrum is 80, 118, and 120 in the JHK peaks,
respectively. Synthetic 2MASS photometry \citep{2003AJ....126.1090C} of the SpeX spectrum yields J-K$_s=2.55$ and J-H$=1.52$, in good agreement with the 2MASS photometry.  \citet{2003PASP..115..389V}'s investigation of the SpeX calibrations suggests the uncertainty in synthetic colors is a few percent. We have also computed synthetic MKO photometry \citep{Tokunaga:2002fj,Tokunaga:2005vn} and found that $J_{2M} - J_{MKO} = 0.12$, $H_{2M} - H_{MKO} = -0.08$ and $K_{s,2M} - K_{MKO} = 0.06$, consistent with the values for L dwarfs found by \citet{Stephens:2004rt}, and leading to predicted value of $J_{MKO}=15.48$, $H_{MKO}=14.08$, and $K_{MKO} = 12.99$. Except where specifcally indicated, we use 2MASS photometry through the remainder of this paper.

\begin{figure}
\plotone{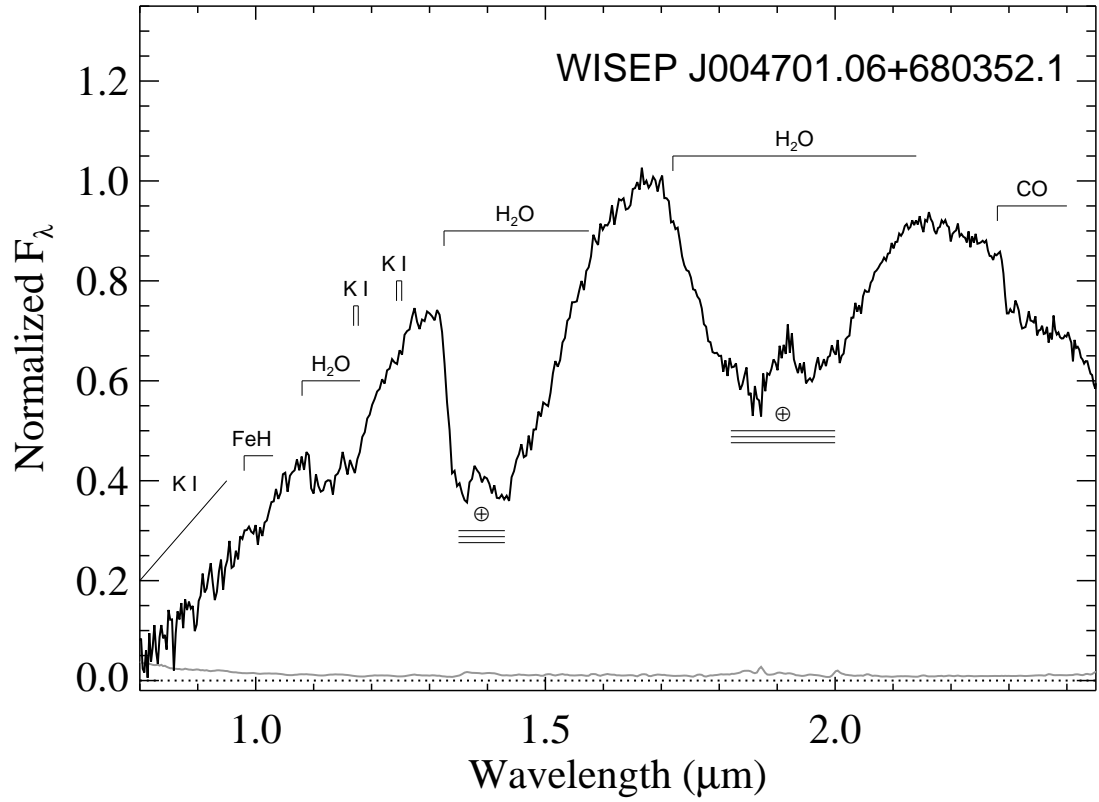}
\caption{SpeX low-resolution near-infrared spectrum of W0047+6803. The uncertainties are shown as a dotted line.  Unlike most L dwarfs, the spectrum peaks in the H band. The steam (H$_2$O) and CO bands are strong, but the atomic lines and FeH appear weak.
\label{fig-spectrum}}
\end{figure}

The overall appearance of the near-infrared spectrum (Figure~\ref{fig-spectrum}) is L dwarf-like but the source is redder than normal L dwarfs, which reach an average J-K$_s \approx 1.82 \pm 0.07$ at spectral type L6 \citep{2010AJ....139.1808S} and whose spectra peak in the J-band.  The red color and deep water absorption bands distinguish W0047+6803 from the typical L dwarf sequence.  For example, the H$_2$O-J and H$_2$O-H indices defined by \citet{2006ApJ...637.1067B} have values of 0.59 and 0.60, corresponding to spectral type T0 \citep{2006ApJ...637.1067B,2007ApJ...659..655B}, but there is no observable methane absorption. We compare W0047+6803 to other L dwarfs in Section~\ref{secLcomp}.

\subsection{Additional Imaging}

To confirm the proper motion suggested by the 2MASS and WISE data, J and K-band images were obtained on UT Date 6 August 2011 with the ASTROCAM \citep{Fischer:2003fj,2004AJ....127.2948V}  at the US Naval Observatory (USNO) 1.55 m Strand Astrometric Reflector. We fit the observed ASTROCAM positions of background reference stars to the 2MASS positions. We fit the 2MASS-USNO observations and find the proper motion relative to the background stars: $\mu_\alpha \cos \delta = 0.381 \pm 0.012$ arcsec yr$^{-1}$ and $\mu_\delta = -0.212 \pm 0.012$ arcsec yr$^{-1}$.  

W0047+6803 was also observed on UT Date 27 August 2011 with the 1.55 m Strand Astrometric Reflector at the Flagstaff Station of the US Naval Observatory using the E2V 2048x4100 CCD Camera and a $z$-filter. The data were reduced using standard techniques and calibrated to the known SDSS magnitude ($z=17.12$ AB mag, \citealt{Aihara:2011fk}) of the L5 dwarf \object{2MASS J01443536-0716142} \citep{2003AJ....125..343L}, observed immediately beforehand. The observed magnitude of W0047+6803 is $z = 18.72 \pm  0.05$ AB mag.  The resulting color, $z-J=3.1 \pm 0.1$, is redder than normal L dwarfs ($z-J=2.8$, \citealt{2010AJ....139.1808S}). Excluding unresolved L/T systems, the only L dwarf measured to be redder in this color seems to be the URL L5 dwarf \object{2MASS J09175418+6028065} ($z-J=3.48 \pm 0.32$, \citealt{Geisler:2011fk}.) 

To estimate the apparent bolometric magnitude, we make use of the similarity between W0047+6803 and \object{2MASSW J2244316+204343} (2M2244+2043) which will be demonstrated in Section~\ref{secLcomp}.  We integrate the W0047+6803 near-infrared spectrum and WISE photometric measurements supplemented by the observed 2M2244+2043 optical \citep{2008ApJ...689.1295K} and mid-infrared \citep{Stephens:2009qy} spectra multiplied by a factor of 2.44, corresponding to their relative K$_s$ brightnesses. The spectral energy distribution beyond these observed ranges 
($\lambda < 0.6$ \micron, $\lambda > 14$ \micron) was extended with blackbody relations, but contributes very little flux. We find an apparent bolometric magnitude of $16.37 \pm 0.08$ (BC$_K = 3.32$), with the dominant uncertainty due to the fact between   
2.5 and 5.2 \micron~ we have no spectra, but only the W1 and W2 broadband photometry.  The uncertainty was determined by using different \citet{Madhusudhan:2011yq} models over the missing range, but requiring agreement with the WISE photometry. 

\section{Discussion\label{sec-discussion}}

\subsection{Comparison to known red L dwarfs\label{secLcomp}}

One approach to classifying URLs is to compare their spectra to ordinary L dwarf standards over a limited wavelength range, to avoid the effect of their extreme color over large wavelength ranges. \citet{Kirkpatrick:2011lr}, for example, classify a number of URLs as ``L9 pec" based on comparing their spectra in the J-band region to L dwarf standards.  \object{WISEPA J164715.59+563208.2} is classified ``L9 pec," although with J-K$_s = 2.20 \pm 0.10$ it is bluer than W0047+6803. We compare the J-band (1.10--1.35 \micron), H-band (1.50--1.80 \micron), and K-band (2.00--2.30 \micron)  to the L dwarf near-infrared standards at the SpeX Prism Spectral Libraries.  In all three bands, the best match when minimizing the least squares differences is the L7 dwarf \object{2MASSI J0103320+193536} \citep{2000AJ....120..447K,Cruz:2004fr}. However, the L8 standard \object{2MASSW J1632291+190441} \citep{1999ApJ...519..802K,2007ApJ...659..655B} is better than the other L7s \object{DENIS-P J0205.4-1159} (D0205-1159, \citealt{1999AJ....118.2466M,1999ApJ...519..802K,2006ApJ...637.1067B})
or \object{2MASS J09153413+0422045} \citep{2008AJ....136.1290R,2007ApJ...659..655B}. These comparisons suggest a spectral type of ``L7 pec" or ``L7.5 pec" for W0047+6803.

\begin{figure}
\plotone{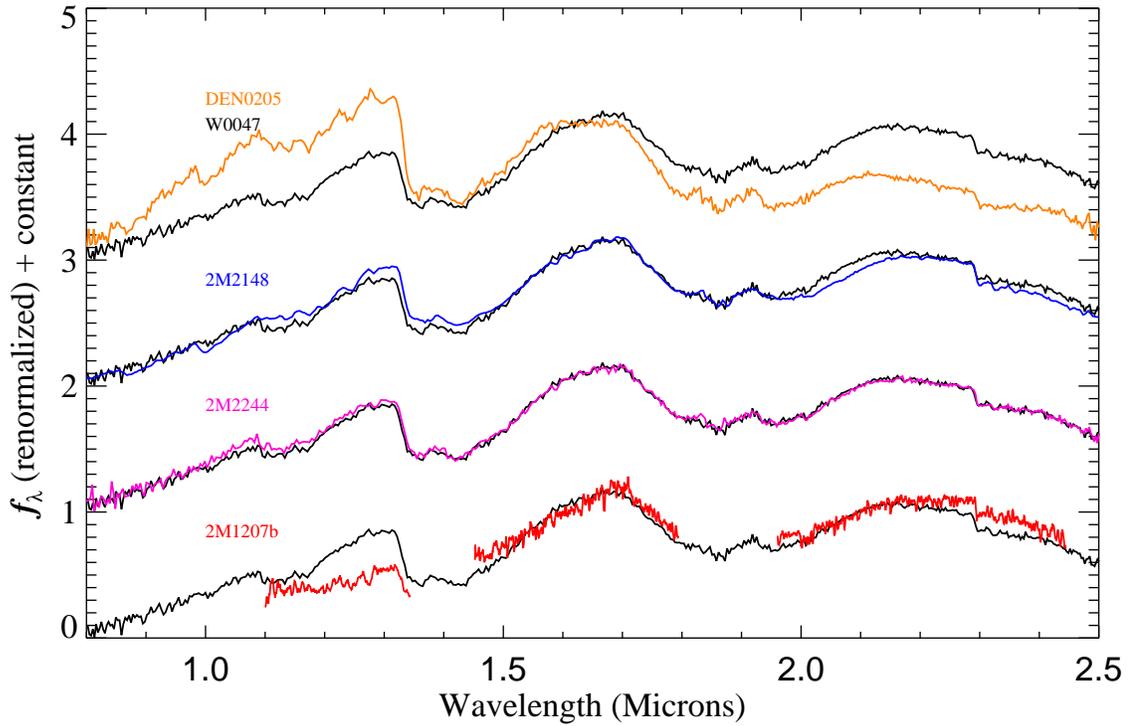}
\caption{W0047+6803 compared to published spectra of the L7 dwarf D0205-1159 \citep{2006ApJ...637.1067B}, the red L dwarfs 2M2148+4003 and 2M2244+2043 \citep{Looper:2008lr}, and the young planetary mass object 2M1207b \citep{2010A&A...517A..76P}. The spectra have been normalized to one for the H-band (1.6 \micron) peak.
\label{fig-redLs}}
\end{figure}

Another approach is to directly compare the spectrum to already classified URL dwarfs. The reddest  (J-K$_s > 2.3$) known field L dwarfs are listed in Table~\ref{tab2}, with all photometry in the 2MASS system. In Figure~\ref{fig-redLs}, we compare the W0047+6803 near-infrared spectrum to the (optical) standard L7 dwarf D0205-1159, the field URL dwarfs \object{2MASS J21481628+4003593} (2M2148+4003, \citealt{Looper:2008lr}) and \object{2MASSW J2244316+204343} (\citealt{dahn,Looper:2008lr}), and the young planetary-mass L-type brown dwarf 2M1207b \citep{2004A&A...425L..29C,2010A&A...517A..76P}. W0047+6803 and 2M2244+2043 ($J-K_s=2.45 \pm 0.16$) are very similar, with W0047+6803 only slightly redder.  We therefore adopt \citet{Looper:2008lr}'s near-infrared classification of ``L7.5 pec" for W0047+6803 as well.  2M1207b is a poor match: It is redder (J-K$_s=3.1 \pm 0.2$, \citealt{edgeon1207}), with deeper water absorption bands and a sharper peak at H-band. The brown dwarf \object{2MASS J03552337+1133437} \citep{2009AJ....137.3345C}
has been classified in the red as an L$5\gamma$ (the $\gamma$ indicates very low surface gravity), and has J-K$_s = 2.52 \pm 0.03$, but no near-infrared spectrum has been published.  If the earlier and later type peculiar URL dwarfs form a sequence with W0047+6803, then L7.5 pec may mark the point at which the URL sequence reaches its maximum in J-K$_s$; on the other hand, 2M1207b suggests that redder colors are possible.  

\begin{figure}
\plotone{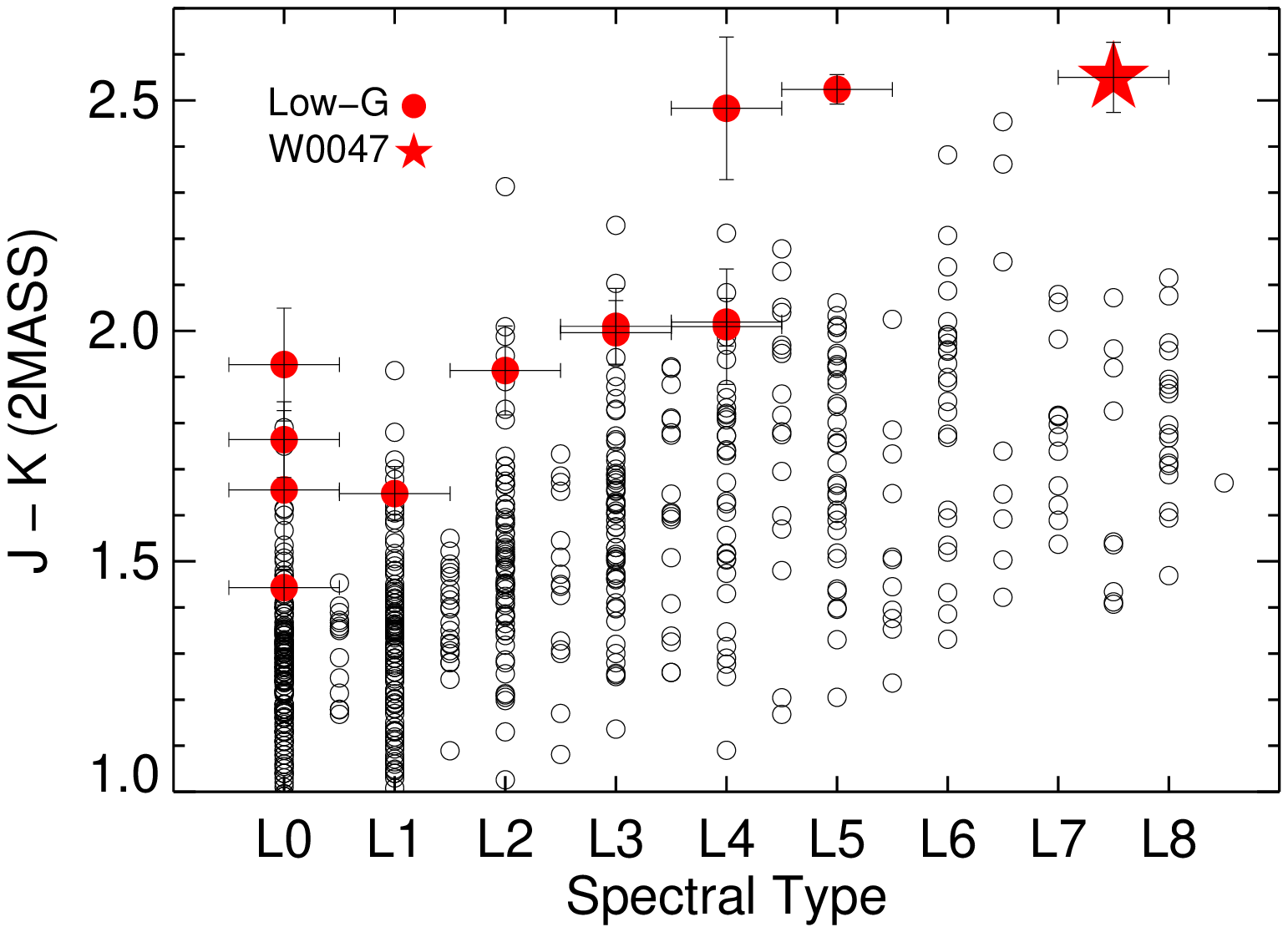}
\plotone{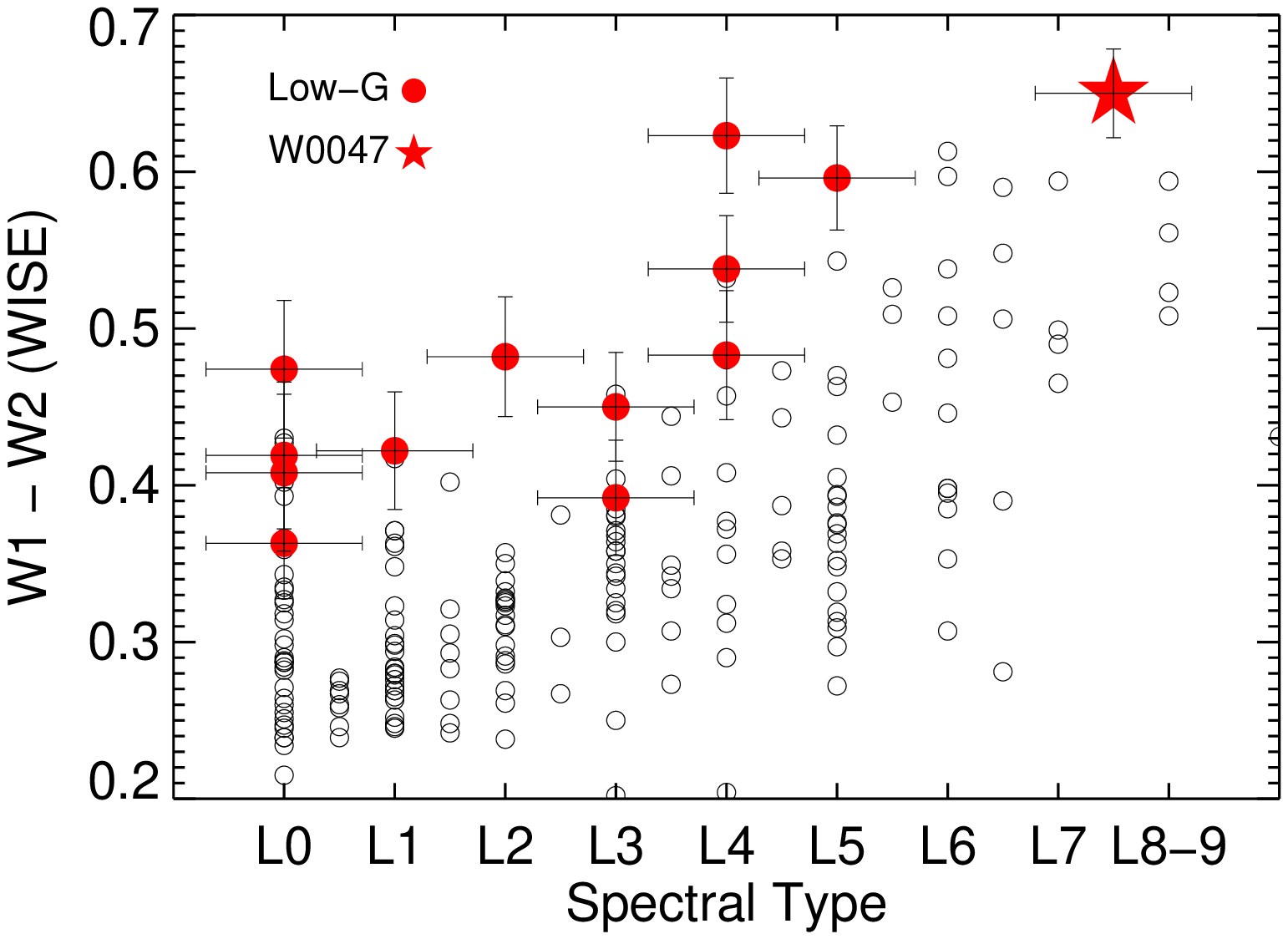}
\caption{The colors of W0047+6803 (star symbol) compared to ordinary field L dwarfs (open circles) and low surface-gravity L dwarfs (solid circles). 
\label{fig-colors}}
\end{figure}

The 2MASS and WISE colors of W0047+6803 are compared in Figure~\ref{fig-colors} to a sample of ordinary L dwarfs as well as a sample of field low surface gravity L dwarfs \citep{2009AJ....137.3345C}, possibly members of local moving groups. These youngest field brown dwarfs are typically URLs, but not all all URLs are low gravity \citep{2009AJ....137.3345C}.  W0047+6803 appears to lie on the extension of the URL (low-gravity) sequence, but other factors than low gravity may be responsible for this.

\begin{figure}
\plotone{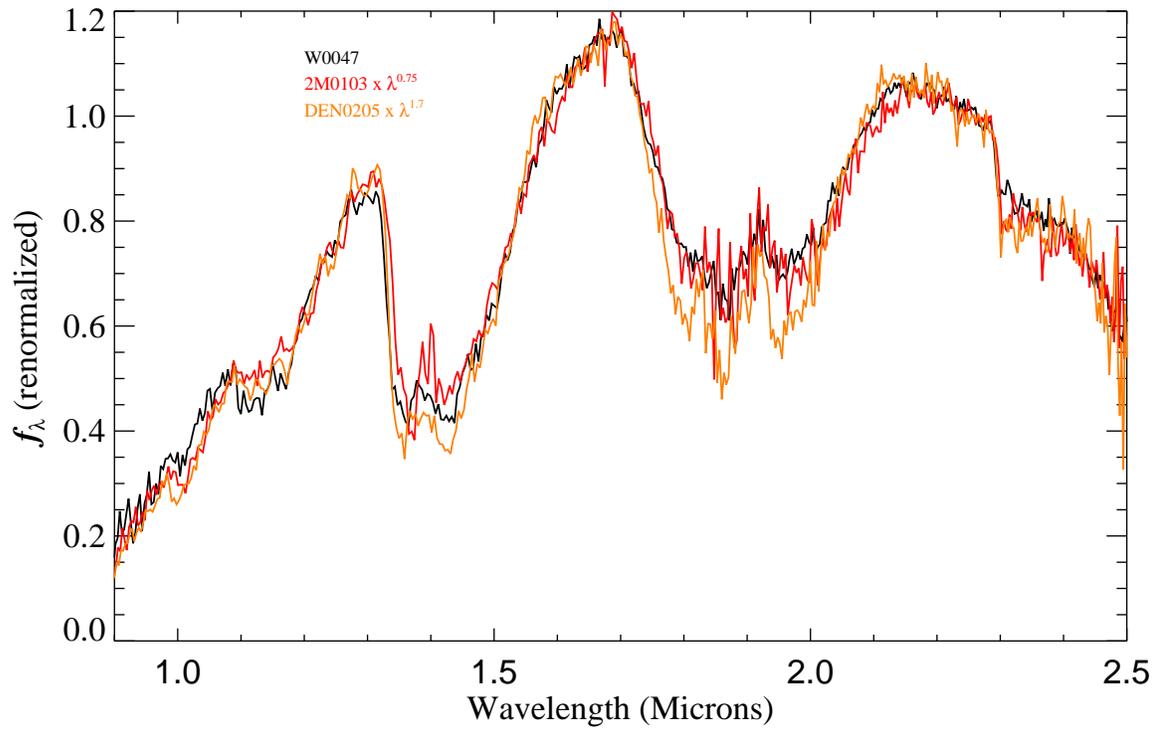}
\caption{W0047+6803 compared to the L7 dwarf D0205-1159 reddened by multiplying by $\lambda^{1.7}$ and the intrinsically redder L7 2M0103+1935 \citep{Cruz:2004fr} reddened by multiplying by $\lambda^{0.75}$. The spectra have been normalized to one at 1.6 \micron.
\label{fig-reddened}}
\end{figure}

Low-gravity or not, it seems likely that URL dwarfs have unusually thick condensate, dusty clouds. \citet{McLean:2003lr} found weak atomic K lines in the near-infrared and the peak in the spectral energy distribution at H-band for 2M2244+2043. W0047+6803 shows the latter trait but our spectrum lacks the resolution necessary to measure the atomic lines. \citet{Zapatero-Osorio:2005qy} found 2M2244+2043 to be strongly polarized at I-band suggesting high levels of dust, although \citet{Goldman:2009mz} did not confirm this measurement.  The Spitzer mid-infrared spectrum of 2M2244+2043 is also consistent with a dusty atmosphere \citep{Stephens:2009qy}, with it and other URLs showing a silicate absorption feature at 9-11 \micron \citep{Looper:2008lr}. \citet{2008ApJ...689.1295K} find 2M2244+2043 is a normal L6.5 with no lithium in the far-red, but speculate that it may nevertheless be a young, low surface-gravity dwarf. Although the H-band appears more peaked on both W0047+6803 and 2M2244+2043, a trait that is linked to low surface gravity for late-M and L dwarfs \citep{Lucas:2001lr,2007ApJ...657..511A,Allers:2010lr,Bihain:2010fk}, the effect for these two objects is not as dramatic as it is for the 2M1207b spectrum shown in Figure~\ref{fig-redLs}. In Figure~\ref{fig-reddened}, we plot a comparison of W0047+6803 to two other L dwarfs, where the other L dwarfs have been reddened by multiplying by a power law $\left(\frac{\lambda}{1.625 \mu \rm{m}}\right)^{\alpha}$.  For the L7 D0205-1159 (J-K$_s=1.6$), $\alpha=1.7$, and for the intrinsically redder L7 \object{2MASSI J0103320+193536}  (J-K$_s=2.14$, \citealt{2000AJ....120..447K,Cruz:2004fr}), $\alpha = 0.75$.  (Note that 2M0103+1935 is classified as a low-surface gravity L6$\beta$ by \citealt{faherty}).
The results suggest that the ``peakiness" in H-band for W0047+6803 is consistent with the overall reddening of the spectrum. \citet{2009AJ....137.3345C} and \citet{2010ApJS..190..100K} argue that a subset of the URL dwarfs are old.  \citet{Allers:2010lr} also show a distinction between low-gravity and older URL dwarfs.  We conclude that W0047+6803's age or surface gravity cannot be constrained by its unusual near-infrared properties or the available data, and that W0047+6803 and 2M2244+2043 may be old.

It appears unlikely that the URL phenomenon is due to binaries. Neither 2M2244+2043 \citep{2003AJ....125.3302G} nor 2M2148+4003 \citep{Reid:2008uq} were resolved by Hubble Space Telescope imaging, and no combination of ordinary bluer L or T dwarfs could produce such red 2MASS colors.

\subsection{Distance and Motion}

Lacking a trigonometric parallax for W0047+6803, we can only estimate distance by comparison to similar L dwarfs of known distance. Unfortunately, none of the late-type URL dwarfs in Table~\ref{tab2} have published trigonometric parallaxes. We can proceed by adopting a spectral type of L7.5 and using empirical relationships between spectral type and absolute magnitude for field L dwarfs, but recognizing that the unusual color will lead to inconsistent results. Using the MKO absolute magnitude relations of \citet{faherty}, the estimated distance would be 12 pc (K) to 18 pc (J); using the mean 2MASS absolute magnitudes calculated by \citet{Dupuy:2012lr} yields 10 to 14 pc. The \citet{Kirkpatrick:2011lr} relationship for W2 suggests 9.7 pc.   \citet{faherty} has found that low surface gravity dwarfs are typically $\sim 0.5$ magnitudes underluminous for their type in the near-infrared bands, which would reduce the estimated distances to 10 (K) to 14 (J) pc.  A direct comparison to \citet{faherty}'s parallax for 2M0103+1935 yields a distance of $12.9 \pm 2.1$ (K$_s$) or $15.5 \pm 2.6$  (J) pc.  On the other hand, 2M2244+2043 is classified L6.5 in the optical, and using this value increases the distance to 13 to 20 pc  \citep{faherty} or 12 to 18 pc \citep{Dupuy:2012lr}.  Use of any of these relations has a $\sim 15$\% uncertainty for typical L dwarfs. The overall range of 10 to 20 pc in plausible distances corresponds to a range of 21 to 41 km s$^{-1}$, which does not usefully constrain the age \citep{2009AJ....137....1F}. W0047+6803 has been added to both the CFHT \citep{Dupuy:2012lr} and USNO \citep{2004AJ....127.2948V} parallax programs, so this distance uncertainty may be resolved in a few years. 

The closer set of distances suggest an interesting possibility. If W0047+6803 is at a distance of $\sim 9-10$pc, its tangential motion would be consistent with membership of the young $\beta$ Pic Association \citep{2001ApJ...562L..87Z,2006A&A...460..695T}. This association is usually thought to be $\sim 12$ Myr old, although \citet{Macdonald:2010lr} argue for an age of $\sim 40$ Myr.  If W0047+6803 were a member, and if the younger age is correct, it would likely be below the deuterium-burning limit. The predicted radial velocity for group membership is $\sim -9$ km s$^{-1}$. However, the low luminosity implied by this close distance would require a effective temperature considerably cooler than that of ordinary field L7 dwarfs. We next discuss the temperature of W0047+6803 using theoretical models.    

\subsection{Comparison to theoretical models\label{sec-theory}}

\begin{figure}
\plotone{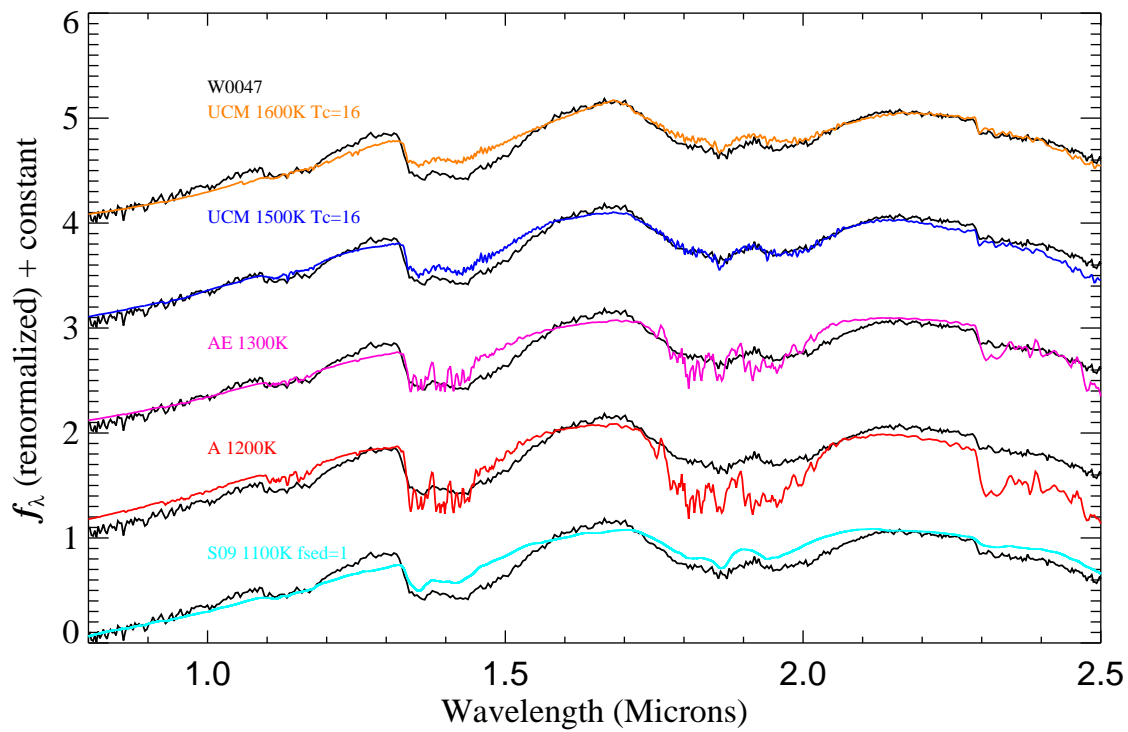}
\caption{Best fitting models to W0047+6803. Top two: the UCM (Unified Cloudy Models) from an updated version of \citet{2002ApJ...575..264T} with $T_{cr}=1600$K.  Middle two: A and AE models from \citet{Madhusudhan:2011yq}. Bottom: \citet{Stephens:2009qy} model with $f_{sed}=1$ and $K_{zz} = 10^6$. All require thicker condensate dust clouds than in normal field L dwarfs.  
\label{fig-theory}}
\end{figure}

A number of models that have been applied to ordinary L dwarfs can be adjusted to produce URL-like spectra. Each of these treats clouds in different ways. For a comparative analysis of different L dwarf cloud models with test cases, see \citet{2008MNRAS.391.1854H}. In Figure~\ref{fig-theory}, we present fits of the observed W0047+6803 spectrum to various model predictions. In each case, we picked the best fit by minimizing the least squares differences between the observed and model spectra. (Since the distance is unknown, the normalization was allowed to take on any value.)  At bottom, we show the fit for 2M2244+6803 presented by \citet{Stephens:2009qy}, which we adopt given the similarity of the two spectra, but it should be noted that this fit is based on both the near-IR and mid-IR spectra. In this model system \citep{Ackerman:2001fj,Marley:2002fk}, the parameter $f_{sed}$ describes the rainout of condensates and $K_{zz}$ describes vertical mixing. Matching 2M2244+6803/W0047+6803 then requires $f_{sed}=1$ (rather than $f_{sed} =2-3$ for normal L dwarfs) and $K_{zz} = 10^6$ (rather than $10^4$.) This model fit is quite cool, with $T_{eff}=1100$K and a normal surface gravity $\log g = 5.0$.  In this context, W0047+6803 may be understood as a dusty object with little rainout and strong vertical mixing that prevented it from following the normal transition to an early-to-mid T dwarf.  

\begin{figure}
\plotone{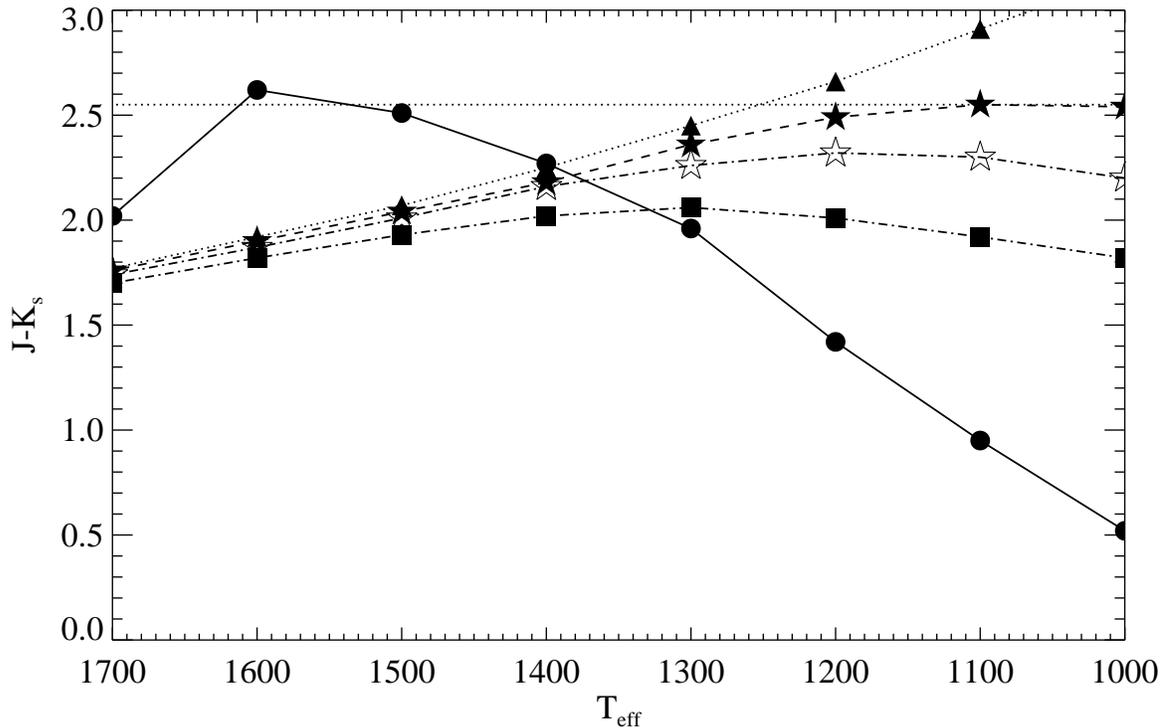}
\caption{Predicted J-K$_s$ colors as a function of effective temperature. W0047's observed color is shown as the horizontal dotted line.   The UCM models with $T_{cr}=1600$K are shown as solid filled circles. Also plotted are the \citet{Madhusudhan:2011yq} A model with $\log g = 4.0$ (solid triangle) and
AE models with $\log g = 4.0$ (solid star), $\log g = 4.5$ (open star), and $\log g = 5.0.$ (solid squares). \label{fig-tcolors}}
\end{figure}

In the middle of Figure~\ref{fig-theory}, we show the best fits using the \citet{Madhusudhan:2011yq} synthetic spectra. We consider models with 1700K $\ge T_{eff} > 1000$K (in steps of 100K.)  In these models, normal L dwarfs are described by the ``E" class, in which the condensate clouds have a definite top and bottom. In the ``AE" models, the cloud decks extend higher in the atmosphere, with an exponential falloff, and in the ``A" models, the cloud decks have no top. For the AE models, we consider $5.0 \ge \log g \ge 3.5$ (in steps of 0.5); for A models only $\log g =4.0$ models are available. These fits have $T_{eff} = 1200$K (A) and $T_{eff} = 1300$K (AE).  Within the AE model grid, high gravities do not produce 2MASS colors as red as W0047+6803. The AE best fits are for low surface gravity ($\log g=4.0$), but this result must be viewed with caution given the higher surface gravity of the \citet{Stephens:2009qy} model and the comparisons in Section~\ref{secLcomp}.  Both model families  agree however that the effective temperature of such red objects are comparable to ordinary field T dwarfs \citep{2004AJ....127.3516G}. This is also shown in Figure~\ref{fig-tcolors}, which shows that red colors are characteristic of low temperatures in these models.  Neither the \citet{Madhusudhan:2011yq} E models ($\log g =4.0$) nor \citet{2006ApJ...640.1063B} E models ($\log g =5.0$) reach near-infrared colors red enough to fit W0047+6803 satisfactorily.

The top two model spectra of Figure~\ref{fig-theory} show ``Unified Cloudy Models" (UCM) of \citet{2002ApJ...575..264T}. In these models, at a critical temperature, $T_{cr}$, the dust particles grow rapidly and rain out of the atmosphere, so the dust cloud deck is found at $T_{cr} < T < T_{cond}$. For normal L and T dwarfs, $T_{cr} = 1800$K \citep{2002ApJ...575..264T,2003ApJ...585L.151T,2005ApJ...621.1033T}.  We find that W0047+6803 may be fit with $T_{cr} = 1600$K. The model grid includes $1800$K$\ge T_{eff} \ge 1100$K in steps of 100K, but have been computed only for low surface gravity ($\log g =4.0$) and solar abundance. The effective temperatures in these fits, $T_{eff} = 1500-1600$K are considerably warmer than in the the other models presented above, and in agreement with the temperatures deduced for optical L6.5 dwarfs \citep{2004AJ....127.3516G}.  As seen in Figure~\ref{fig-tcolors}, these models predict a turn to a bluer T-dwarf-like colors for lower temperatures even with dust.  Besides the three sets of models we have presented, it should be noted that \citet{Looper:2008lr} have shown GAIA models (albeit for early type URLs) where high metallicity create URL-like spectra.  These models also favor warmer effective temperatures, like the UCM models, but they do not have low gravity.  

In summary, we have two different theoretical scenarios for the temperature of W0047+6803 and the other reddest URL L dwarfs in Table~\ref{tab2}. On the one hand, they could be have effective temperatures typical of T dwarfs ($T_{eff} \approx 1100-1300$K), and have developed  thicker clouds rather than the usual clearing. Alternatively, they may have temperatures similar to their L5-L7 dwarf counterparts, with $T_{eff} \approx 1500-1600$K), again with thicker clouds. (Presumably reality may also be in between these extremes.) Perhaps the fact that 2M2244+2043's optical spectrum is a normal L6.5 type favors the warmer temperatures. We note that if the radii were the same, the absolute bolometric magnitude of a $T_{eff} = 1200$K brown dwarf would be one magnitude fainter  than for $T_{eff}=1500$K, or equivalently, the hotter solution is 60\% further away. A parallax offers the possibility of resolving this issue, if the radius, age, or surface gravity of W0047+6803 can be independently estimated. High metallicity may be a cause of the thicker clouds but has not been explicitly included in the models used here.  

\section{Conclusion}

W0047+6803 is a bright, nearby, unusually red L dwarf. Its observed spectrum can be interpreted as due to an extremely dusty atmosphere with thick condensate clouds.  Similar characteristics occur in low-surface gravity brown dwarfs and planetary mass objects, although we cannot constrain the age or surface gravity of W0047+6803, and there is mounting evidence that some URL dwarfs are old. Overall, the lack of ``peakiness" compared to 2M1207b and other confirmed young L dwarfs suggest W0047+6803 is not unusually young, but until the cause of the unusually red spectrum is fully understood young ages cannot be rejected. As a bright, nearby object, W0047+6803 is well suited for additional studies of extremely dusty substellar atmospheres, and should lead to a better understanding of the URL class, the characteristics of young gas-giant planets, and the L/T transition in field brown dwarfs. If it proves to be an older URL, then it may help reveal the relative roles of metallicity, surface gravity, and possibly other parameters in creating thick, dusty condensate clouds.   

\acknowledgments

We thank Adam Burgasser, Adam Burrows, Davy Kirkpatrick, Sandy Leggett, Dagny Looper,  Nikku Madhusudhan, Mark Marley, Ian McLean, Jenny Patience, Didier Saumon and Takashi Tsuji, as well as  their collaborators, for sharing their spectra in electronic form.  We thank the anonymous referee for helpful suggestions.

This publication makes use of data products from the Wide-field Infrared Survey Explorer, which is a joint project of the University of California, Los Angeles, and the Jet Propulsion Laboratory/California Institute of Technology, funded by the National Aeronautics and Space Administration. This publication makes use of data products from the Two Micron All Sky Survey, which is a joint project of the University of Massachusetts and the Infrared Processing and Analysis Center/California Institute of Technology, funded by the National Aeronautics and Space Administration and the National Science Foundation. This research has made use of the NASA/ IPAC Infrared Science Archive, which is operated by the Jet Propulsion Laboratory, California Institute of Technology, under contract with NASA. This research has made use of the VizieR catalogue access tool, CDS, Strasbourg, France. This research has made use of the SIMBAD database, operated at CDS, Strasbourg, France. The Digitized Sky Surveys were produced at the Space Telescope Science Institute under U.S. Government grant NAG W-2166. The images of these surveys are based on photographic data obtained using the Oschin Schmidt Telescope on Palomar Mountain and the UK Schmidt Telescope. The plates were processed into the present compressed digital form with the permission of these institutions. 
The National Geographic Society - Palomar Observatory Sky Atlas (POSS-I) was made by the California Institute of Technology with grants from the National Geographic Society.
The Second Palomar Observatory Sky Survey (POSS-II) was made by the California Institute of Technology with funds from the National Science Foundation, the National Geographic Society, the Sloan Foundation, the Samuel Oschin Foundation, and the Eastman Kodak Corporation. This research made use of APLpy, an open-source plotting package for Python hosted at \url{http://aplpy.github.com}. This research has benefitted from the M, L, and T dwarf compendium housed at \url{DwarfArchives.org} and maintained by Chris Gelino, Davy Kirkpatrick, and Adam Burgasser. This research has benefitted from the SpeX Prism Spectral Libraries, maintained by Adam Burgasser at \url{http://www.browndwarfs.org/spexprism}

\bibliographystyle{apj}

\begin{deluxetable}{lc}
\tablewidth{0pc}
\tabletypesize{\footnotesize}
\tablenum{1} \label{tab1}
\tablecaption{WISEP J004701.06+680352.1 }
\tablehead{
\colhead{Parameter} & 
\colhead{W0047+6803} }
\startdata
WISE RA(J2000) & 00 47 01.06 \\
WISE Dec (J2000) & +68 03 52.1\\
WISE Epoch & 2010.09 \\
$z$ [mag] & $18.72 \pm 0.05$ \\
2MASS J [mag] &  $15.60 \pm 0.07$ \\
2MASS H [mag] & $13.97 \pm 0.04$ \\
2MASS K$_s$ [mag] & $13.05 \pm 0.03$ \\
WISE W1 [mag] & $11.90 \pm 0.02$ \\ 
WISE W2 [mag] & $11.25 \pm 0.02$ \\
WISE W3 [mag] & $10.14 \pm 0.06$ \\
WISE W4 [mag] & $>8.42$ \\
$\mu_\alpha \cos \delta$ (mas/yr) & $381 \pm 12$ \\
$\mu_\delta$ (mas/yr) & $-212 \pm 12$ \\
Sp Type (Near-IR) & L$7.5$ pec\\
$m_{bol} $ & $16.37 \pm 0.08$ \\
\enddata
\end{deluxetable}

\begin{deluxetable}{lclll}
\tablewidth{0pc}
\tabletypesize{\footnotesize}
\tablenum{2} \label{tab2}
\tablecaption{The reddest field L dwarfs }
\tablehead{
\colhead{Object} &
\colhead{J-K$_s$} &
\colhead{Sp (Opt)} & 
\colhead{Sp (nIR)} & 
\colhead{Reference}}
\startdata
2MASS J21481628+4003593 & $2.38 \pm 0.04$ & L6 & L6.5 pec &  1 \\
2MASSW J2244316+204343 & $2.45 \pm 0.16$ & L6.5 & L7.5 pec & 1,2,3 \\
2MASS J16154255+4953211 & $2.48 \pm 0.16$ & \nodata & L5? & 4\\
2MASS J03552337+1133437 & $2.52 \pm 0.03$ & L5$\gamma$ & \nodata & 5 \\
WISEP J004701.06+680352.1 & $2.55 \pm 0.08$ & \nodata & L7.5 pec & 6 \\
\enddata
\tablerefs{1. \citet{Looper:2008lr} 2.  \citet{dahn} 
3. \citet{2008ApJ...689.1295K} 4. \citet{Geisler:2011fk}
5. \citet{2008AJ....136.1290R} 6. This paper }
\end{deluxetable}

%

\end{document}